\documentclass[review]{elsarticle}

\usepackage{lineno,hyperref}
\usepackage{amsfonts,amsmath,amssymb}
\usepackage{xcolor}

\begin{document}

\begin{frontmatter}

\title{Giant caloric effects close to \emph{any} critical end point}

\author[unesprioclaro]{Lucas Squillante}
\author[unesprioclaro]{Isys F. Mello}
\author[unespilhasolteira]{A. C. Seridonio}


\author[unesprioclaro]{Mariano de Souza\corref{mycorrespondingauthor}}
\cortext[mycorrespondingauthor]{Corresponding author}
\ead{mariano.souza@unesp.br}

\address[unesprioclaro]{S\~ao Paulo State University (Unesp), IGCE - Physics Department, Rio Claro - SP, Brazil}
\address[unespilhasolteira]{S\~ao Paulo State University (Unesp), Department of Physics and Chemistry, Ilha Solteira - SP, Brazil}

\begin{abstract}
The electrocaloric effect (ECE), i.e., the reversible temperature change due to the adiabatic variation of the electric field, is of great interest due to its potential technological applications in refrigeration. Based on entropy arguments, we present a new framework to attain giant ECE. Our findings are fourfold: \emph{i}) we employ the recently-proposed electric Gr\"uneisen parameter $\Gamma_E$ to quantify the ECE and discuss its advantages over the existing so-called electrocaloric strength;
\emph{ii}) prediction of giant caloric effects \emph{close} to \emph{any} critical end point; \emph{iii}) proposal of potential key-ingredients to enhance the ECE; \emph{iv}) demonstration of $\Gamma_E$ as a proper parameter to probe quantum ferroelectricity in connection with the celebrated Barrett's formula. Our findings enable us to interpret the recently-reported large ECE at room-temperature in oxide multilayer capacitors [Nature \textbf{575}, 468 (2019)], paving thus the way for new venues in the field. \newline\newline
\end{abstract}

\begin{keyword}
A. multilayers, A. electronic materials, D. crystal structure, D. dielectric properties, D. ferroelectricity.
\end{keyword}

\end{frontmatter}


\section*{Introduction}

Nowadays, there is an increasing scientific interest regarding new cooling devices that can be both efficient and more environmentally friendly, see, e.g., Refs.\,\cite{Moya,Scott} and references cited therein. Much efforts have been made to develop such devices employing, for instance, both magnetocaloric (MCE) \cite{Liu} and electrocaloric (ECE) effects \cite{Mathur}, i.e., the cooling as a consequence of the adiabatic removal of magnetic or electric field, respectively. When dealing with magnetocaloric devices, one of the limitations is the requirement, in many cases, to have a superconducting refrigerated permanent magnet, reducing thus the cost-benefit of employing such devices in technological applications. The main advantage of electrocaloric devices in turn is highlighted by the fact that only an adiabatic electric field sweep is needed for cooling the system. More recently, the ECE became even a more attractive topic because of the discovery of large electrocaloric effect in oxide multilayer capacitors \cite{Mathur}.
\begin{figure}[!h]
\centering
\includegraphics[clip,width=\columnwidth]{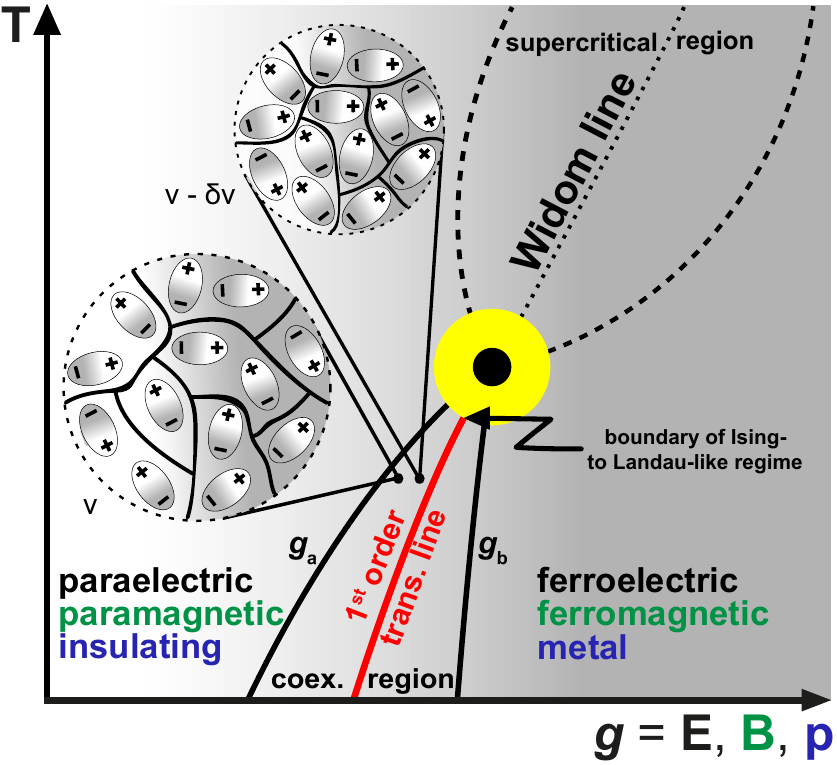}
\caption{\footnotesize \textbf{The phases coexistence region, the first-order transition line and the finite-temperature critical end point.} Schematic phase diagram temperature $T$ \emph{versus} tuning parameter $g$ showing the coexistence region, the first-order transition line, the spinodal lines $g_a$ and $g_b$, the Widom line embedded in the supercritical region, and the finite-$T$ critical end point (black bullet). The various phases with their corresponding tuning parameters, namely electric field $E$, magnetic field $B$, and pressure $p$, are also shown. Within the coexistence region, two distinct electric-dipoles configurations are schematically shown, corresponding to two distinct volumes, namely $v$ and $v - \delta v$.  Upon increasing $p$, the electric dipolar interaction is enhanced and the entropy is also increased as a direct consequence of approaching the first-order transition line and the critical end point. The region between the electric-dipoles represents the non-polarized state.
The yellow region indicates the Landau-like regime \cite{Zacharias}, being the border of the change from Ising- to Landau-like regime also indicated. More details in the main text.}
\label{Fig-1}
\end{figure}
Considering the ECE in real systems, it is natural to look for its optimization aiming to increase the benefits in potential practical applications, as well as aiming to lower the cost. However, the mechanism for attaining maximized ECE, as well as the appropriated conditions for such, remains elusive. Here, based on entropy arguments, we propose potential key-ingredients to achieve giant ECE. Figure \ref{Fig-1} depicts a schematic phase diagram, namely temperature as a function of the tuning parameter, highlighting the merge of the coexistence region into the second-order critical end point. The latter is key in understanding the enhancement of the ECE in the vicinity of a critical end point, as we discuss into more details in the following. It is clear that in order to have a more pronounced ECE, the entropy of the system must be enhanced. Since the entropy is an extensive physical quantity, by summing up all contributions to the total entropy, we can achieve  higher entropy values and thus a more expressive ECE \cite{Liu-APL}. In order to estimate the ECE, the so-called EC strength, namely $\Delta T$/$\Delta E$, is usually employed, where $T$ is the temperature and $E$ the electric field. In this framework, systems that present a high temperature variation with relatively low $E$ adiabatic sweeps are considered good candidates for large ECE. Nonetheless, by employing solely the EC strength it is not straightforward to perform a systematic investigation as a function, for instance, of $T$ and $E$ for different systems aiming to maximize the ECE. This is particularly true  specially because the EC strength does not take into account the starting temperature of the adiabatic process, preventing thus a proper comparison between the ECE for different systems. We use the electric Gr\"uneisen parameter $\Gamma_E$, proposed in Ref.\,\cite{submittedprl}, to estimate the ECE taking the entropy variations in respect to $T$ and $E$ into account, as well as the starting temperature.
Note that $\Gamma_E$ naturally embodies the non-linear effects of the ECE since the adiabatic $T$ derivative of $E$ is taken into account, being thus distinct from the case of the usually employed $\Delta T/\Delta E$ in the EC strength approach \cite{littlewood}. We show that $\Gamma_E$ is not only more appropriate, when compared with the usual EC strength, to analyze the entropy sources for the ECE, but also it can be considered as a \emph{smoking-gun} for predicting maximized ECE in real systems. We also discuss electrostriction effects and demonstrate their relation with the structural instability entropy contribution, which in turn is associated with one of the contributions responsible for the enhancement of the ECE. Furthermore, we generalize the Gr\"uneisen parameter for the various existent caloric effects and discuss $\Gamma_E$ in connection with Barrett's \cite{Barret} formula for ferroelectric quantum critical systems.\newline

\section*{Results}
\subsection*{The electric Gr\"uneisen parameter and the electrocaloric effect}

The well-known magnetocaloric effect can be quantified by the so-called magnetic Gr\"uneisen parameter \cite{prbmce}. Analogously, $\Gamma_E$ quantifies the  ECE \cite{submittedprl}:
\begin{equation}
\Gamma_E = \frac{-\left(\frac{\partial P}{\partial T}\right)_E}{c_E} = -\frac{1}{T}\frac{\left(\frac{\partial S}{\partial E}\right)_T}{\left(\frac{\partial S}{\partial T}\right)_E} = \frac{1}{T}\left(\frac{\partial T}{\partial E}\right)_S.
\label{ECEequation}
\end{equation}
where $c_E$, $P$, and $S$ refer, respectively, to the specific heat at constant $E$, the electric polarization and the entropy. It turns out that $\Gamma_E$ is also connected with the material dielectric constant $\varepsilon$ by the relation $\Gamma_E = \frac{E}{c_E}\left(-\frac{\partial\varepsilon}{\partial T}\right)_E$ \cite{submittedprl}. In relaxor-type ferroelectrics, upon tuning the system close to the critical end point by applying external pressure \cite{Gesi}, the dielectric response can be dramatically altered since the Coulomb repulsion between charges is enhanced and so does the entropy and $E$ required to polarize the system, cf.\,Fig.\ref{Fig-1}. Recently, we have proposed that the relaxation time is entropy-dependent for \emph{any} system close to the critical end point \cite{submittedprl}. We have shown that close to the first order transition line of \emph{any} phases coexistence region the relaxation time is significantly enhanced and we are faced with a Griffiths-like phase. As a consequence, we infer here that \emph{any} caloric effect will be enhanced close to the first order transition line and critical end point. This is one of the main results of this work.\newline

\subsection*{Giant caloric effects close to \emph{any} critical end point}  Upon approaching a critical end point (Fig.\,\ref{Fig-1}), an intrinsic entropy accumulation takes place due to the phases coexistence and presence of the critical fluctuations \cite{Gene,Barto,Souza2015, EPJ,PRL2007}. Considering the Maxwell-relation $\left(\frac{\partial P}{\partial T}\right)_{E} = \left(\frac{\partial S}{\partial E}\right)_{T}$ \cite{Zhong,Pirc},  an enhancement of $\left(\frac{\partial S}{\partial E}\right)_{T}$ near the critical point is also reflected in an increase of $\left(\frac{\partial P}{\partial T}\right)_{E}$. Hence, it is straightforward to infer that a giant ECE takes place near the $E$-induced finite-$T$ critical point \cite{Kutnjak}, since such effect is quantified by $\Gamma_E$, which also depends on $\left(\frac{\partial P}{\partial T}\right)_{E}$, cf.\,Eq.\,\ref{ECEequation}. Interestingly, in Ref.\,\cite{Dressel} the authors report on $\epsilon$ measurements for the spin-liquid candidate $\kappa$-(BEDT-TTF)$_2$Cu$_2$(CN)$_3$,  being $\epsilon = \varepsilon/\varepsilon_0$ and $\varepsilon_0$ the vacuum permittivity.   There, a significant increase of $\epsilon$ close to the critical end point is observed. By carefully analyzing such $\epsilon$ experimental results \cite{Dressel}, an estimate of $\Gamma_E$ taking into account the temperature derivative of the electric polarization (or dielectric constant) shows that $\Gamma_E$ can reach giant values. Indeed, for $p =$ 1.45\,kbar, using a frequency of 7.5\,kHz, at $T \approx 10$\,K, i.e., upon approaching the critical point,  $\Gamma_E \approx 10^{-5}$\,m/V. For PbSc$_{0.5}$Ta$_{0.5}$O$_3$, a large ECE near the critical point around room-$T$ was recently reported \cite{Mathur}. Making use of the data reported in Ref.\,\cite{Mathur} regarding the driving $E$ and the effective $T$ change $\Delta T$, we estimate the ECE to be $\Gamma_E \approx 10^{-9}$\,m/V. The value of $\Gamma_E$ calculated for PbSc$_{0.5}$Ta$_{0.5}$O$_3$ is associated with a temperature variation $\Delta T \approx 5.5$\,K \cite{Mathur}, which is considered as a large ECE in the literature. Thus, although the numerical value of $\Gamma_E$ itself for this system is relatively low, it is indeed associated with a large ECE at room-temperature. The estimated value of $\Gamma_E$ for $\kappa$-(BEDT-TTF)$_2$Cu$_2$(CN)$_3$ under the optimal conditions of frequency and pressure is much larger than that estimated for PbSc$_{0.5}$Ta$_{0.5}$O$_3$. Such a result is suggestive that the ECE in $\kappa$-(BEDT-TTF)$_2$Cu$_2$(CN)$_3$ should be more pronounced.  The relatively high compressibility inherent to molecular conductors together with their layered structure \cite{Barto,PRL2007} can be considered as possible key factors for such a large ECE. The $\Gamma_E$ values for both systems here considered were computed under the conditions corresponding to points in their phase diagrams, which are close to the first-order transition line and critical point. These are real examples that illustrate the main result of this work regarding a giant ECE near the critical end point employing $\Gamma_E$. Now, we generalize the Gr\"uneisen parameter for \emph{any} external parameter tuned adiabatically. Following Eq.\,\ref{ECEequation}, we write a generalized expression for the Gr\"uneisen parameter considering any external tuning parameter $g = B, E, p$, being $B$ the external magnetic field and $p$ the pressure (Fig.\,\ref{Fig-1}):
\begin{equation}
\Gamma_g = \frac{1}{T_s}\left(\frac{\partial T}{\partial g}\right)_S,
\label{ECE-geral}
\end{equation}
where $T_s$ is the starting temperature of the adiabatic process.
Hence, following Eq.\,\ref{ECE-geral}, our analysis can be extended to \emph{any} other physical scenarios with distinct tuning parameters.  Table \ref{Table-Gru} summarizes the various caloric effects here discussed.
\begin{table}[!h]
\centering
\begin{tabular}{|c|c|}
\hline\hline
Maxwell-Relation & Gr\"uneisen Parameter  \\ \hline\hline
$\left(\frac{\partial S}{\partial p}\right)_T = -\left(\frac{\partial v}{\partial T}\right)_p$ & $\Gamma = -\frac{1}{Tv_m}\frac{\left(\frac{\partial S}{\partial p}\right)_T}{\left(\frac{\partial S}{\partial T}\right)_p} = \frac{1}{T v_m}\left(\frac{\partial T}{\partial p}\right)_S$     (BCE)               \\ \hline
$\left(\frac{\partial S}{\partial B}\right)_T = \left(\frac{\partial M}{\partial T}\right)_B$ & $\Gamma_{mag} = -\frac{1}{T}\frac{\left(\frac{\partial S}{\partial B}\right)_T}{\left(\frac{\partial S}{\partial T}\right)_B} = \frac{1}{T}\left(\frac{\partial T}{\partial B}\right)_S$ (MCE)                     \\ \hline
$\left(\frac{\partial S}{\partial E}\right)_T = \left(\frac{\partial P}{\partial T}\right)_E$ & $\Gamma_E = -\frac{1}{T}\frac{\left(\frac{\partial S}{\partial E}\right)_T}{\left(\frac{\partial S}{\partial T}\right)_E} = \frac{1}{T}\left(\frac{\partial T}{\partial E}\right)_S$          (ECE)           \\ \hline
$\left(\frac{\partial S}{\partial P}\right)_T = -\left(\frac{\partial E}{\partial T}\right)_P$ & $\Gamma_P = -\frac{1}{T}\frac{\left(\frac{\partial S}{\partial P}\right)_T}{\left(\frac{\partial S}{\partial T}\right)_P} = \frac{1}{T}\left(\frac{\partial T}{\partial P}\right)_S$              (PCE)        \\ \hline
\end{tabular}
\caption{\footnotesize
The thermodynamic ($\Gamma$), magnetic ($\Gamma_{mag}$), electric ($\Gamma_E$), and polar ($\Gamma_P$) Gr\"uneisen parameters with their corresponding Maxwell-relations. 
BCE, MCE, ECE, and PCE refer, respectively, to the barocaloric, magnetocaloric, electrocaloric, and polarcaloric \cite{Benepe} effects; $v_m$ is the molar volume and $M$ the magnetization. Details in the main text.}
\label{Table-Gru}
\end{table}

\begin{figure}[htb!]
\centering
\includegraphics[clip,width=0.58\columnwidth]{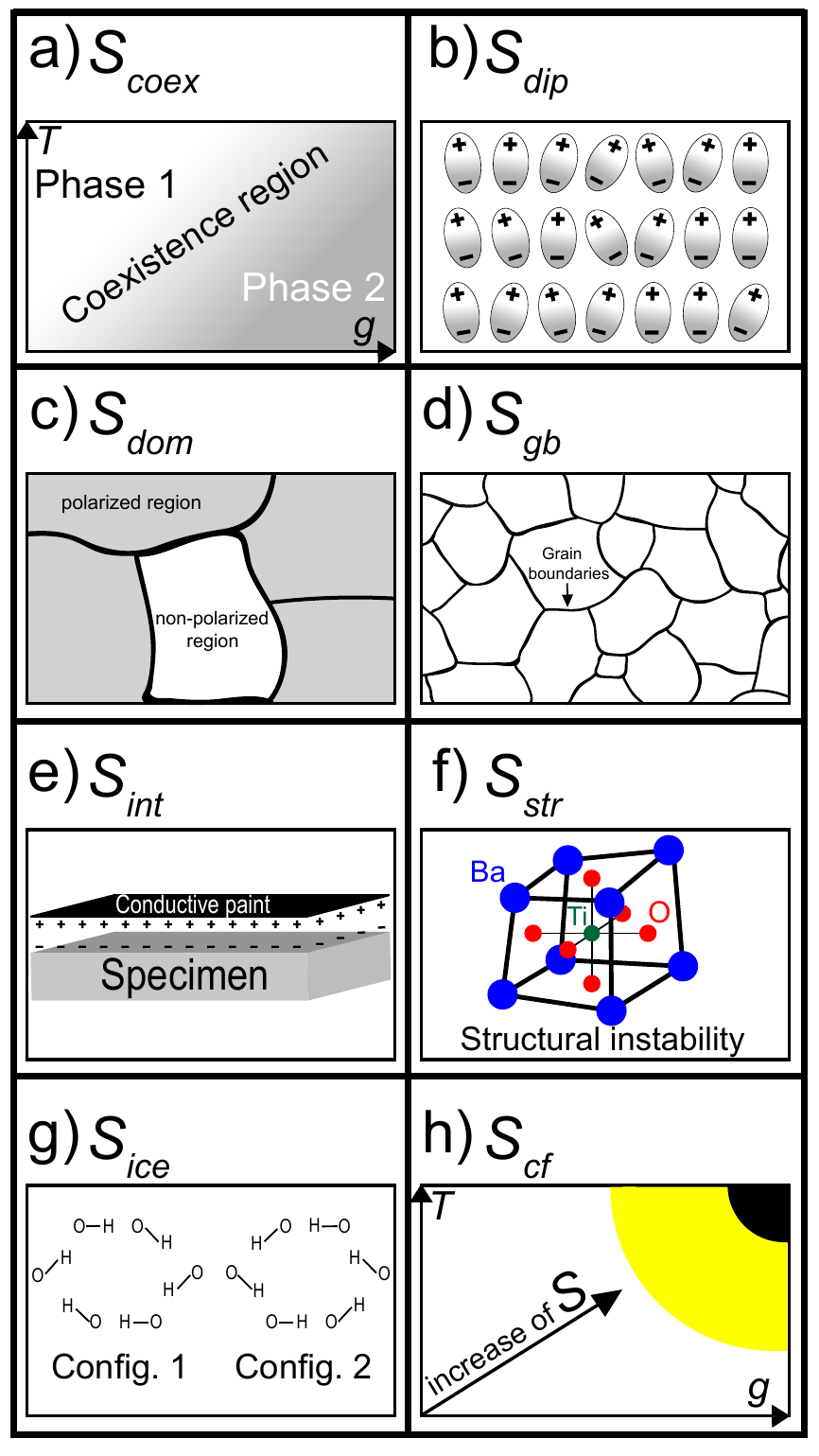}
\caption{\footnotesize \textbf{The various entropy contributions to the total entropy.} Schematic representation of the various contributions to the total entropy. a) Coexistence of two generic phases 1 and 2; b) distinct electric dipolar configurations in a ferroelectric material; c) ferroelectric domains and the coexistence of polarized and non-polarized regions; d) presence of grain-boundaries, usually inherently present in poly-crystalline materials; e) a finite electric polarization existent between the conductive paint and the specimen, which is responsible for the so-called Maxwell-Wagner polarization; f) cartoon of a distorted BaTiO$_3$ BCC unit cell, illustrating the $S_{str}$ contribution to the total entropy; g) residual entropy contribution $S_{ice}$ associated with two distinct molecular configurations, cf. proposed by L. Pauling \cite{linuspauling}, and h) enhancement of the total entropy due to critical fluctuations upon approaching the finite-$T$ critical end point (Fig.\,\ref{Fig-1}). Details in the main text.}
\label{Fig-2}
\end{figure}

Interestingly enough, to the best of our knowledge, although hitherto not properly discussed in the literature, the origin of the Gr\"uneisen parameter lies on Maxwell-relations when considering adiabatic processes \cite{Anna,XMoya}. This is universal and can be extended to all forms of writing the Gr\"uneisen parameters reported in the literature \cite{EPJ,Zhu,Kuchler,SR,prbmce}, cf.\,Table \ref{Table-Gru}.
Since an entropy accumulation inherently takes place near any critical point, the adiabatic variation of the corresponding tuning parameter towards its critical value will enhance its respective caloric effect. For instance, if pressure is removed adiabatically, the barocaloric effect will be pronounced near the critical pressure. This can also be extended for both the magnetocaloric and electrocaloric (polarcaloric) effects. Of course, such an entropy accumulation near the critical point depends on the properties of the system of interest, being the shape of the first-order transition line (Fig.\,\ref{Fig-1}) also crucial.
Regarding relaxor-type ferroelectric systems \cite{Kutnjak}, upon applying external pressure the system can be tuned to approach its critical point \cite{Gesi}. In a simple picture, the entropy can be associated with the degree of disorder into the system, i.e., the number of possible configurations, and thus it is natural to consider that the disorder intrinsic to relaxor-type ferroelectrics is significantly enhanced near the critical point, which in turn is associated with the previously-mentioned entropy accumulation. Our analysis suggests that there are some key-ingredients to maximize the ECE. This is discussed in the following. A possible generalized expression for the various contributions to the entropy of a system candidate to giant ECE reads:
\begin{equation}
S = S_{coex} + S_{dip} + S_{dom} + S_{gb} + S_{int} + S_{str} + S_{ice} + S_{cf},
\label{entropiatotal}
\end{equation}
where $S_{coex}$ refers to the entropy associated with the coexistence region near a critical point; $S_{dip}$ is the entropy contribution originating from distinct dipolar configurations; $S_{dom}$ is associated with the presence of domains in relaxor-type ferroelectrics; $S_{gb}$ is the grain-boundaries entropy contribution inherent to poly-crystalline materials; $S_{int}$ represents the entropy contribution due to interface-effects, e.g., Maxwell-Wagner polarization \cite{maxwellwagner1,maxwellwagner2,maxwellwagner3,NatureMaterials}; $S_{str}$ corresponds to the contribution associated with structural instabilities as it occurs, for instance, in BaTiO$_3$ \cite{Str-Inst}; $S_{ice}$ is associated with the residual entropy of the possible configurations of the molecules in a solid \cite{linuspauling}, and $S_{cf}$ takes into account the entropy contribution from the critical fluctuations in the Ising-regime, cf.\,Fig.\,\ref{Fig-1}. Note that at some extent the enhancement of the ECE is directly associated with the coexistence of both polarized and non-polarized regions. This highlights the relevance of the coexistence region near finite-$T$ critical points \cite{Kutnjak,banani,Debenedetti}. The various contributions to the total entropy (Eq.\,\ref{entropiatotal}) are schematically illustrated in Fig.\,\ref{Fig-2}.
Hence, in order to maximize the ECE a system should incorporate more contributions to the total entropy as possible.
Regarding the interface effect, the authors of Ref.\,\cite{lunkenheimer} discuss its role on giant $\epsilon$ based on Maxwell-Wagner polarization \cite{maxwellwagner1,maxwellwagner2,maxwellwagner3}. Furthermore, in Ref.\,\cite{Serghei2009} the role played by the thickness of the interface $d$ in the dielectric constant $\varepsilon^{*} = (\varepsilon' - i\varepsilon'')$ is given by $L/\varepsilon^*_{m} = [2d/\varepsilon^{*}_i + (L-2d)/\varepsilon^{*}_b$], where $L$ is the specimen length, $\varepsilon^*_{m}$ is the measured dielectric constant, $\varepsilon^{*}_i$ and $\varepsilon^{*}_b$ refer to the interface and bulk dielectric constant, respectively. Note that when $2d \rightarrow L$,  $\varepsilon^*_{m}$ will mostly be due to the interface effects contribution. In other words, when the length of the specimen is comparable to $2d$, the interface contribution will dominate. Considering that $\Gamma_E$ depends on the $T$ derivative of $\varepsilon$, higher values of $\varepsilon$ due to interface effects can also contribute to the enhancement of the ECE, since $S_{int}$ will be enhanced as well. Yet, the ECE is enhanced with the increase of the latent heat of a first-order transition upon approaching the critical point and, as a consequence, with the corresponding associated $S$ variation. Hence, following such arguments, the high entropy accumulation near critical points, due to the critical fluctuations and/or the proximity to the first-order transition line, can give rise to an even more pronounced ECE. In Ref.\,\cite{Mathur} the authors were able to drive the ferroelectric transition to the supercritical regime and thus to access 1.5 times the entropy associated with the latent heat of such a transition, increasing thus the ECE. In general terms, following previous discussions, our findings suggest that the maximization of the ECE is essentially fourfold: \emph{i}) the ECE will be enhanced in the coexistence regime of the phase diagram of any system, being not maximum in the critical end point itself, but near it due to the proposed change from Ising- to Landau-like (mean-field) regime near the critical point \cite{Zacharias}. More specifically, the ECE will be maximized upon approaching the first-order transition line in the region, in which the system changes from Ising- to Landau-regime, cf.\,Fig.\,\ref{Fig-1}; \emph{ii}) the relaxor-like character of ferroelectrics is also important due to the intrinsic enhancement of $S$ inherent to the presence of ferroelectric domains; \emph{iii}) multilayered systems present interface effects, such as the Maxwell-Wagner polarization, being associated with another entropy contribution that can enhance the ECE; \emph{iv}) the structural instabilities contribute to the enhancement of the total entropy and thus to the increase of the ECE. In an ideal situation,  the ECE will be maximized when all contributions to the total entropy, described in Eq.\ref{entropiatotal}, are taken into account. It would be challenging to design a material in this framework.  Yet, a similar discussion as the one here proposed for the maximization of the ECE can also be performed for the magnetocaloric effect. In this regard, the recently reported relatively large magnetocaloric effect in Gd$_3$Ni/Gd$_{65}$Ni$_{35}$ composite microwires constitutes a meaningful example \cite{Phan2020}. The coexistence of magnetic phases of both nanocrystalline Gd$_3$Ni and amorphous Gd$_{65}$Ni$_{35}$, together with the structural disorder introduced by the amorphous phase, lead to an enhancement of the magnetic entropy change of the system and thus to a large magnetocaloric effect.
\newline 

\subsection*{Electrostriction effects}

Following previous discussions, we define the electrostriction coefficient. Using Maxwell-relations \cite{Gene}, the magnetostriction coefficient can be related with the pressure derivative of the magnetization $M$, as follows \cite{Hafner1985}:
\begin{equation}
\left(\dfrac{\partial M}{\partial p}\right)_T = -\frac{1}{v\mu_0}\left(\dfrac{\partial v}{\partial H}\right)_T,
\label{magnetostriction}
\end{equation}
where $\mu_0$ is the vacuum magnetic permeability, $v$ is the volume of the specimen, and $H$ the magnetic field strength. Note that the $M$ variation upon applying pressure is directly associated with the volume change upon applying external magnetic field. Thus, analogously, for the electric case, we define:
\begin{equation}
\left(\dfrac{\partial P}{\partial p}\right)_T = -\frac{\varepsilon_0}{v}\left(\dfrac{\partial v}{\partial D}\right)_T,
\label{electrostriction}
\end{equation}
where $D$ is the modulus of the electric displacement vector.
Note that the electrostriction effect, namely the volume variation upon applying an electric field, is associated with the $p$-dependence of $P$. At this point, it is worth recalling that
the electromechanical strain is given by $x = QP^2$, where $Q$ is the  electrostriction coefficient \cite{Sundar1992,Bell1992}. Thus, since $P^2 \propto E^2$, being $P = \varepsilon_0 \chi E$, we can infer that $x \propto E^2$ and $\Delta T = \frac{1}{2}\frac{\beta T}{c_{E}}P^2$, being $\chi$ the electric susceptibility and $\beta$ a phenomenological parameter, and consequently $x \propto \Delta T$ \cite{Lu2012}. Considering the Gibbs free energy obtained from Landau's theory, we can compute the entropy variation as a function of $Q$ \cite{Lu2012}, namely:
\begin{equation}
\Delta S = -\frac{1}{2}\beta P^2+QP^2\dfrac{\partial \sigma}{\partial T}+s\sigma \dfrac{\partial \sigma}{\partial T} + \alpha\left(\sigma + T \dfrac{\partial \sigma}{\partial T}\right),
\label{isothermal ECE}
\end{equation}
where $s$ refers to the elastic compliance coefficient, $\sigma$ refers to the stress, and $\alpha$ is the thermal expansion coefficient. Equation \ref{isothermal ECE} describes the variation of $S$ associated with structural changes in a system, being thus related with the contribution $S_{str}$ to the total entropy in Eq.\,\ref{entropiatotal}.
\newline
\subsection*{The electric Gr\"uneisen parameter and quantum ferroelectricity}

In a classical paraelectric-to-ferroelectric transition, it is well-known that $\varepsilon$ as a function of $T$ is maximized at the ferroelectric transition temperature, indicating that the system becomes more polarizable.
\begin{figure}[htb!]
\centering
\includegraphics[clip,width=\columnwidth]{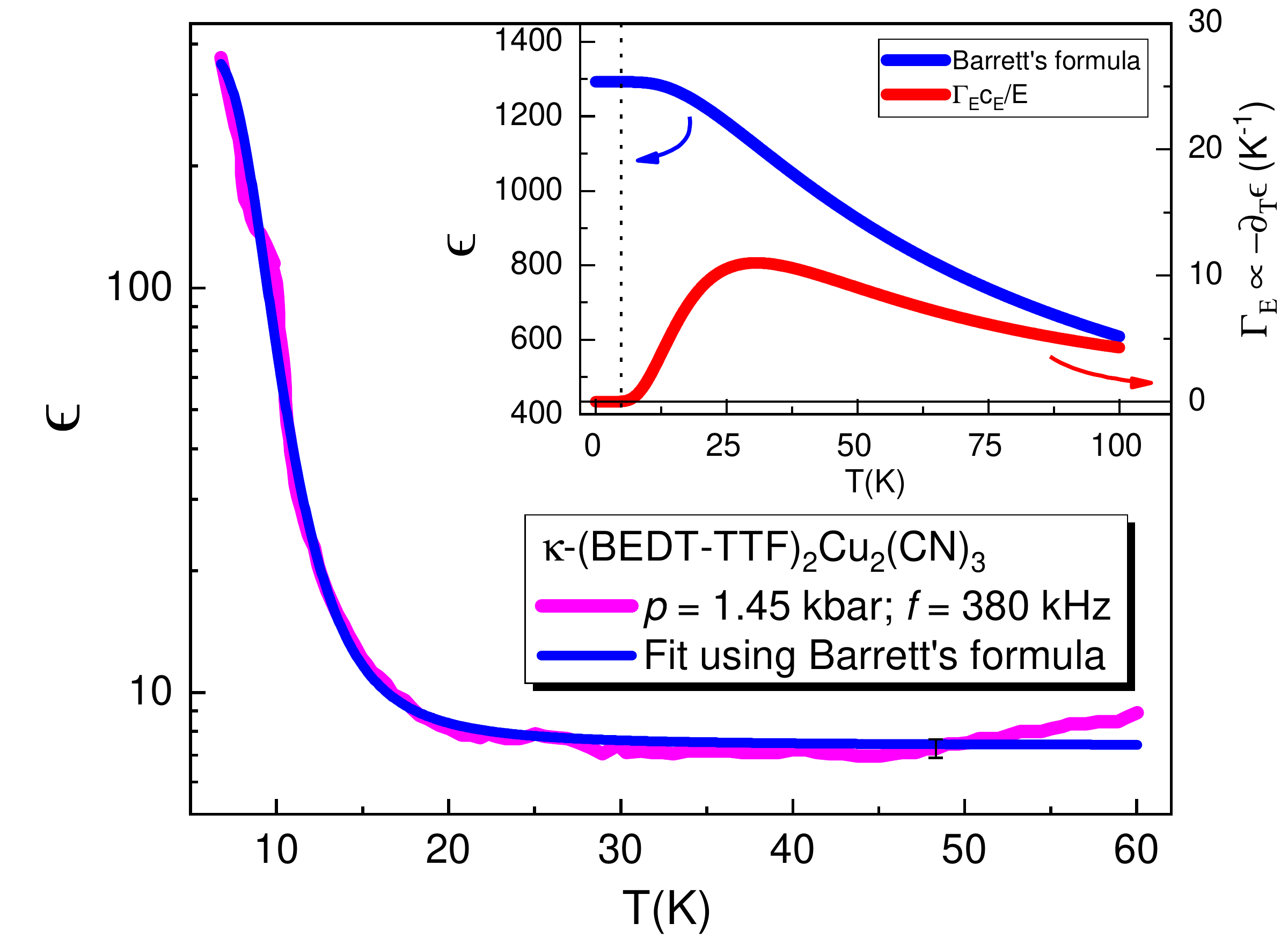}
\caption{\footnotesize \textbf{The dielectric constant of the spin-liquid candidate $\kappa$-(BEDT-TTF)$_2$Cu$_2$(CN)$_3$ and Barrett's formula.} Main panel: dielectric constant $\epsilon$ as a function of temperature $T$ (magenta solid line) for the spin-liquid candidate $\kappa$-(BEDT-TTF)$_2$Cu$_2$(CN)$_3$ under pressure $p$ = 1.45\,kbar, frequency $f$ = 380\,kHz, and fitting of the data set employing Barrett's formula (blue solid line). Experimental data set taken from Ref.\,\cite{Dressel}. The vertical error bar refers to a 5\% error in the data tracing. Inset: Temperature dependence of the dielectric constant $\epsilon$ using Barrett's formula \cite{Barret} (blue solid line) and $\Gamma_E \propto -\partial_T \epsilon$ in terms of Barrett's formula (Eq.\,\ref{Barret}) (red solid line). The  parameters are $A$ = 0, $m$ = (8.4 $\times$ 10$^4$)\,K (using CGS units), $T_1$ = 60\,K, and $T_0$ = $-$35\,K \cite{Barret}. The vertical dashed line represents the temperature in which $\epsilon$ is constant and thus $-\partial_T\epsilon = 0$. Details in the main text.}\label{Fig-3}
\end{figure}
A handful of investigations have been carried out focussing on the behavior of $\varepsilon$ upon applying pressure, aiming to unveil the effect of $p$ on the ferroelectric phase. Several systems reported in the literature present a quantum-phase transition from ferroelectric to a quantum ferroelectric upon pressurization \cite{Saxena,Muller,Samara,Horiuchi}. In such cases, the $T$-dependence of $\epsilon$ follows Barrett's formula \cite{Barret}:
\begin{equation}
\epsilon = A + \frac{m}{\left(\frac{T_1}{2}\right)\coth{\left(\frac{T_1}{2T}\right)}-T_0},
\end{equation}
where $A$ is a non-universal constant, $T_0$ is a parameter related either to the ferroelectric ($T_0 >$ 0) or antiferroelectric ($T_0 <$ 0) effective dipolar interaction, $T_1$ is the temperature below which the quantum fluctuations dominate, and $m = n\mu^{2}/k_B$, where $n$ is the electric-dipole density, $\mu$ refers to the local electric-dipole moment, and $k_B$ is Boltzmann constant. Furthermore, $T_1$ can be defined as the balance between thermal and the vibrational energies associated with a harmonic oscillator in an electric field by the form $T_1 = h\nu/k_B$ \cite{Barret}, where $h$ is Planck's constant and $\nu$ is the oscillation frequency. Essentially, in this analysis the response of the electric-dipoles to an external $E$ is treated in the harmonic approximation. The quantum critical regime is set when the energy associated with such oscillation overcomes the thermal energy. In general terms, the fingerprint of a quantum paraelectric phase is a plateau in $\epsilon$, i.e., the quantum fluctuations affect the ferroelectric phase and instead of presenting a vanishing $\epsilon$ for $T \rightarrow $ 0, as it occurs in classical ferroelectric-type transitions \cite{prb2018}, $\epsilon$ remains constant \cite{Saxena}. It is worth mentioning that such a plateau will be observed only when the ferroelectric transition temperature is close to $T_1$, otherwise the quantum critical fluctuations will not affect $\epsilon$ \cite{Barret}. Since $\Gamma_E$ is related with the temperature derivative of $\epsilon$, we can write it in terms of Barrett's formula:
\begin{equation}
\Gamma_E = \frac{\varepsilon_0 E}{c_E}\frac{m {T_1}^2}{T^2 \left[T_1 \cosh
   \left(\frac{T_1}{2 T}\right)-2
   T_0 \sinh \left(\frac{T_1}{2
   T}\right)\right]^2}.
   \label{Barret}
\end{equation}
\vspace{0.1cm}
Upon analyzing the behavior of $\Gamma_E$ using Barrett's formula (Eq.\,\ref{Barret}), the enhancement of $\Gamma_E$ (left flank) in warming up  marks the temperature in which quantum fluctuations dominate the ferroelectric phase, cf.\,inset of Fig.\,\ref{Fig-3}, which is associated with the maximization of the ECE at $T \sim$ 27\,K. Thus, we propose that a pronounced ECE is expected in quantum ferroelectrics close to the temperature in which the quantum fluctuations dominate the system. Note that $\Gamma_E$ can be considered not only as an indicator to quantify the ECE using entropy arguments, but also it can be seen as an appropriate physical parameter to demonstrate, or even predict, a quantum ferroelectric behavior. Note that the ECE close to a quantum critical point can be even more pronounced when the electric-dipole density $n = N/v$, where $N$ is the number of electric-dipoles, is enhanced upon pressurization. As pressure is increased, $v$ decreases and thus $n$ increases, making thus the ECE more pronounced, cf.\,Eq.\,\ref{Barret}. Thus, employing  Barrett's formula and $\Gamma_E$, we predict that systems presenting a quantum critical ferroelectric ground-state should also show a pronounced ECE.  
Right on the critical end point we have $\partial_g S = 0$ and, as a consequence $\Gamma_E = 0$. 
Yet, it is worth mentioning that the flattening of $\epsilon$ inherently to Barrett's formula at low-$T$ is merely a consequence of the effective electric dipolar interaction and quantified by $T_0$, being an analogous situation also found for magnetic systems \cite{SR2020}. As previously discussed \cite{submittedprl}, the relaxation time is entropy-dependent upon approaching the critical point. As a consequence, the dielectric response of the system at a particular frequency is dramatically affected under pressurization close to the critical region. This particular behavior prevents the analysis of such data set using the Barrett's formula for various $p$ values.  Hence, we stick our analysis to $\epsilon$ measurements reported in Ref.\,\cite{Dressel} for the spin-liquid candidate $\kappa$-(BEDT-TTF)$_2$Cu$_2$(CN)$_3$ under $p$ = 1.45\,kbar and $f$ = 380\,kHz, cf.\,Fig.\,\ref{Fig-3}, i.e., the system is located on the verge of the Mott second-order critical end point. Also, it is worth mentioning that we have limited our analysis until 60\,K in order to rule out possible electrical contacts contribution to $\epsilon$ \cite{Dressel}. Note that  Barrett's formula fits the experimental data set quite well, being $T_1 \approx$ 86\,K obtained in this fitting. Amazingly, the obtained value of $T_1$ in this analysis is in line with the proposal of a finite-$T$ quantum critical behavior of the Mott transition, reported in Refs.\,\cite{Furukawa,kanoda2}. The obtained value of $T_1$ from the fitting indicates that quantum fluctuations associated with the proposed \emph{strange quantum-critical fluid} \cite{Furukawa,kanoda2} set in around such a temperature under the conditions of $p$ and $f$ shown in Fig.\,\ref{Fig-3}. A deeper understanding of this peculiar behavior deserves further investigations. \newline

\section*{Conclusions}
Using the Gr\"uneisen parameter and entropy arguments, we have demonstrated an intrinsic enhancement of caloric effects near \emph{any} finite-$T$ critical end point. Furthermore, we have generalized the Gr\"uneisen parameter for any tuning parameter $g$ strengthening our argument that any caloric effect is more pronounced in the vicinity of a critical end point. We have proposed that $\Gamma_E$ is an appropriate parameter to quantify the ECE in various systems when compared with the usually employed EC strength, since the starting temperature and the entropy contributions to the ECE are taken into account when $\Gamma_E$ is used. Also, using entropy arguments we have proposed potential key-ingredients to enhance the ECE. Yet, we have shown that $\Gamma_E$ is key in analyzing a ferroelectric quantum-critical behavior in connection with Barrett's formula.
Last but not least, it would be challenging to measure caloric effects close to triple points \cite{Debenedetti,Ishchuk}.\newline

\section*{Methods}
All the calculations present in this work are exactly solvable. The software Wolfram Mathematica$^\circledR$ Version 11 was employed to process data. Figures \ref{Fig-1} and \ref{Fig-2} were generated using Adobe Illustrator$^\circledR$; the data set depicted in Fig.\,\ref{Fig-3} was were generated employing the software OriginPro$^\circledR$ Version 2018. The data set shown in the inset of Fig.\,\ref{Fig-2} was generated in Wolfram Mathematica$^\circledR$ Version 11 and plotted with OriginPro$^\circledR$ Version 2018.


\section*{Acknowledgements}
MdeS acknowledges financial support from the S\~ao Paulo Research Foundation - Fapesp (Grants No.\,2011/2250-4, 2017/07845-7, and 2019/24696-0), National Council of Technological and Scientific Development - CNPq (Grants No.\,302498/2017-6), CJR; and TUVSOTE. ACS acknowledges CNPq (Grants No.\,305668/2018-8). This work was partially granted by Coordena\c c\~ao de Aperfei\c coamento de Pessoal de N\'ivel Superior - Brazil (Capes) - Finance Code 001 (Ph.D. fellowship of LS and IFM).

\section*{Author contributions}
MdeS and LS wrote the paper with contributions from ACS and IFM. LS carried out the calculations and generated the figures. All authors revised the manuscript. MdeS conceived and supervised the project.


\end{document}